\documentclass[12pt]{article}

\usepackage{amssymb}
\usepackage{amsthm}

\def\beq{\begin{eqnarray}}
\def\eeq{\end{eqnarray}}
\def\beqs{\begin{eqnarray*}}
\def\eeqs{\end{eqnarray*}}

\newcommand{{\SD}}{\rm SD}

\newcommand{\lan}{\langle}
\newcommand{\rrr}{\rangle}

\newcommand{\be}{\begin{equation}}
\newcommand{\ee}{\end{equation}}
\newcommand{\T}{\mbox{Tr}\> }
\def\centeron#1#2{{\setbox0=\hbox{#1}\setbox1=\hbox{#2}\ifdim
\wd1>\wd0\kern.5\wd1\kern-.5\wd0\fi
\copy0\kern-.5\wd0\kern-.5\wd1\copy1\ifdim\wd0>\wd1
\kern.5\wd0\kern-.5\wd1\fi}}
\def\ltap{\;\centeron{\raise.35ex\hbox{$<$}}{\lower.65ex\hbox{$\sim$}}\;}
\def\gtap{\;\centeron{\raise.35ex\hbox{$>$}}{\lower.65ex\hbox{$\sim$}}\;}

\title{Non-Stationary Measurements of the Chiral Magnetic Effect}

\author{V.I.Shevchenko\thanks{Vladimir.Shevchenko@cern.ch}}

\date{National Research Centre "Kurchatov Institute" \\
ac. Kurchatova sq., 1, Moscow 123182 Russia}

\begin{document}

\maketitle

\begin{abstract}

We discuss Chiral Magnetic Effect from quantum theory of measurements point of view for non-stationary measurements. The effect of anisotropy for fluctuations of electric currents in magnetic field is addressed within this framework. It is shown that the anisotropy caused by nonzero axial chemical potential is indistinguishable from the one caused by finite lifetime of the magnetic field and in all cases it is related to abelian triangle anomaly.   Possible P-odd effects for central collisions (where the magnetic field and hence Chiral Magnetic Effect is absent) are discussed in this context.

\end{abstract}






\section{Introduction}
\label{}

The behavior of matter in magnetic fields stays of interest since the discovery of magnetism and
plenty of outstanding results in condensed matter physics have been obtained on this way.
In recent years considerable attention is attracted to a particular case of chiral fermions in external magnetic field. The examples range from gapless quasiparticle excitations like the ones in graphene to almost massless quarks deconfined for the very short time during heavy-ion collision. While the magnetic field in the former case is applied (and artificially varied) by the experimenter, the latter system is always under the influence of magnetic field  for non-central collisions created by the debris of the colliding ions. The magnitude of this field is fixed by geometry and kinematics of the problem. An important part of heavy ion experimental programs of RHIC at BNL and of LHC at CERN is devoted to the studies of possible physical effects caused by this strong magnetic fields \cite{Voloshin1} - \cite{Wang}.

The most interesting effect discussed in this context in the so called Chiral Magnetic Effect (CME) \cite{Kharzeev1} - \cite{Fukushima3}. From theoretical side, it corresponds to the fact that chirally asymmetric medium of charged massless fermions under external magnetic field conducts electric current given by the following expression:
\be
{\bf J} = \lan \bar\psi {\bf\gamma} \psi \rrr = \frac{e{\bf B}}{2\pi^2}\mu_5
\label{q1}
\ee
where $\mu_5 = (\mu_R - \mu_L)/2$ is the axial chemical potential and $\mu_L, \mu_R$ are chemical potentials for right and left-handed chiral fermions, respectively. The expression (\ref{q1}) can be derived in many complementary ways and from theoretical side is a robust result. It has got support from the lattice \cite{Buividovich1} - \cite{Buividovich4}. On the other hand, the question about clear experimental manifestations of CME is far from being simple. Perhaps the most important reason for that is an obvious discrepancy between stationary character of (\ref{q1}) and highly non-stationary dynamics of heavy-ion collisions. In particular, the magnetic field $B$ of high enough magnitude exists for $~ 0.1 \div 0.2 $ Fm/c and decays with time in a very fast way.

As is advocated in \cite{Shevchenko1,Shevchenko2} the crucial feature of the CME which gives chance for the corresponding effects to be experimentally observable is classicalization of some degrees of freedom. The necessity for that can be roughly explained as follows. At classical level, originally there is no nonzero axial chemical potential in the problem. Quantum-mechanically one could have some superposition of states with different values of $\mu_5$, for a given collision event. However there is no quantum mechanical current (\ref{q1}) in such state, $\lan {\bf J} \rrr = 0$, because state vectors with positive and negative $\mu_5$ have equal weights. To get nonzero measured current (\ref{q1}) in a particular event, this superposition must be projected to a state with definite and nonzero $\mu_5$. This projection takes place in the course of measurement, i.e. interaction of the quantum system under consideration with the classical measuring device. Taking into account that typical heavy ion collision process is characterized by huge energy density and particle multiplicity it is natural to assume that the role of such measuring device is played by the medium itself. All that has clear links to Color Glass Condensate idea \cite{mcl} and even earlier studies of the subject \cite{Pomeranchuk}.

There exist various theoretical frameworks taking into account quantum-to-classical transition. Quantum measurement theory is among well studied ones. The basic idea is to couple some artificial system ("the detector") with the quantum field in question and study the response of the former.
There are in general two groups of factors, having effect on the detector's response: external conditions (fields, temperature etc) and geometric form of the detector's world-line embedded in bulk space-time. In most cases discussed in the literature one's main interest is to compare inertial detector under some external factor in Minkowski  space-time with the response of non-inertial detector. The best known example is the celebrated Unruh result \cite{Unruh} about correspondence between inertial detector in thermal bath and non-inertial detector at constant acceleration. Of much interest is application of this theory in condensed matter context \cite{Zubkov}. In the present paper (like in \cite{Shevchenko2}) the attitude is different: the detector is supposed to be at rest and attention is focussed on dependence of the detector's response on external conditions.

In \cite{Shevchenko2} the finite temperature case is considered. In the present paper we deal with the chiral fermions at finite chemical potential. The effects of finiteness of the measurement interval are discussed. In the last section the problem of P-odd signatures at central collisions is also briefly addressed.

\section{Finite chemical potentials}

For the reader's convenience let us remind basic steps of quantum measurement procedure (\cite{Braginsky,Birrel}, see also \cite{Bessa} in the context of vector current measurements). The Hamiltonian describing field-detector interaction in our case reads:
\be
\Delta H = \int\limits_{-\infty}^{\infty} f(\tau) d\tau \> m(\tau)  \bar\psi(x(\tau)) \Gamma \psi(x(\tau))
\label{detector}
\ee
Here $x(\tau)$ parameterizes the detector's world-line, $\tau$ is the proper time along it, $m(\tau)$ - internal quantum variable (monopole momentum) of the detector whose evolution in $\tau$ is described by the standard two-level Hamiltonian with the levels $E_0$ and $E_1$, $E_1 - E_0 = \omega > 0$. The Lorentz structure of the coupling is fixed by the matrix $\Gamma$, in what follows we concentrate on the vector case, $\Gamma = n_\mu \gamma^\mu$. The dimensionless "window function" $f(\tau)$ encodes the fact that any realistic measurement takes finite time (which we denote $\lambda$ in this paper), so the window function has the important property $f(\tau \lesssim \lambda) \approx 1$, $f(\tau \gg \lambda) \to 0$. By convention the measurement window is located around the point $\tau = 0$.  The interested reader is refereed to \cite{Padm} for detailed discussion of the finite measurement time effects.

An amplitude for the detector to "click" is given by
\be
{\cal A} = i \int\limits_{-\infty}^{\infty} f(\tau)  d\tau \> \lan 1 | m(\tau) | 0 \rrr \cdot \lan \Omega | j(x(\tau)) | \Omega_0 \rrr
\label{y7}
\ee
where $j(x(\tau)) = \bar\psi(x(\tau)) \Gamma \psi(x(\tau))$ and $| \Omega_0 \rrr$ stays for initial state of the field sub-system, while $| \Omega \rrr$ represents final (after the measurement) state.
The probability is proportional to $|{\cal A}|^2$, and the corresponding response function reads:
\be
{\cal F}(\omega) =  \int\limits_{-\infty}^{\infty} f(\tau) d\tau \int\limits_{-\infty}^{\infty} f(\tau')  d\tau' \> e^{-i\omega (\tau - \tau')} \cdot   G^+(\tau - \tau')
\label{or}
\ee
where
\be
G^+(\tau - \tau') = \lan \Omega_0 | j (x(\tau)) j (x(\tau')) | \Omega_0 \rrr
\label{wi}
\ee

For infinite measurement time, i.e. in the limit $f(\tau) \equiv 1$ one is interested in detector's excitation rate in unit time. It is determined by the power spectrum of the corresponding Wightman function:
\be
\dot{{\cal F}}(\omega) = \int\limits_{-\infty}^{\infty}  d s \>  e^{-i\omega s} \>  G^+(s)
\label{eri}
\ee
In general case one has to consider original expression (\ref{or}).

We analyze (\ref{or}) for the case of chiral fermions at finite density and magnetic field in this section. The corresponding Green's functions were studied in many papers, see \cite{Chodos, Shuryak}. To be self-contained we collect some relevant expressions in the Appendix.

The chemical potentials $\mu_L$ and $\mu_R$ stay for left and right chiral components, respectively. The uniform magnetic field $B$ is chosen in the $z$-direction. Since the detector is at rest, we need Green's function $G_{L,R}(x,x')$ for the argument $(x-x')=(t,0,0,z)$ (with subsequent limit $z\to 0$).  The indices $L$ and $R$ stay for "left" and "right" and the latter one should not be confused with "retarded". The exact expressions have the following form:
$$
G_L(x,x') = \lan \psi_L(x) {\bar\psi}_L(x') \rrr =
$$
\be = \frac{1-\gamma^5}{2} \left[ G^{(0)}(\mu_L,t,z)  + \frac{\gamma^1 \gamma^2 eB}{16\pi^2}  G^{(1)}(\mu_L,t,z) + {\cal O}(B^2) \right]
\label{i1}
\ee
and analogously
$$
G_R(x,x') = \lan \psi_R(x) {\bar\psi}_R(x') \rrr =
$$
\be = \frac{1+\gamma^5}{2} \left[ G^{(0)}(\mu_R,t,z)  + \frac{\gamma^1 \gamma^2 eB}{16\pi^2}  G^{(1)}(\mu_R,t,z) + {\cal O}(B^2) \right]
\label{i2}
\ee
where
$$
8\pi^2 (t^2 - z^2)^2 z^2 \> G^{(0)}(\mu,t,z) =
$$
$$
= \left[ \vphantom{\frac12} e^{-i\mu z} (t+z) \left( (tz + z^2)(i+\mu(t-z))(\gamma_0 - \gamma_3) + i \gamma_3 (t^2 - z^2) \right) - \right.
$$
\be
\left. e^{i\mu z} (t-z) \left((tz - z^2)(i+\mu(t+z))(\gamma_0 + \gamma_3) + i \gamma_3 (t^2 - z^2) \right)
 \vphantom{\frac12} \right]
\label{free}
\ee
\be
G^{(1)}(\mu,t,z) = - \frac{1}{t^2 - z^2} \left[ e^{-i\mu z} (\gamma_0 - \gamma_3)(t+z) + e^{i\mu z} (\gamma_0 + \gamma_3)(t-z) \right]
\label{b}
\ee
It is important that Lorentz structure of ${\cal O}(B^2)$ terms contains only $\gamma_0$ and $\gamma_3$ matrices as for the free term (and hence they give no contribution to anomaly and/or asymmetry we are looking for). Notice absence of a factor $\exp(-i\mu t)$ in both expressions (\ref{free}), (\ref{b}), which corresponds to the fact that there is no energy gap. The poles on the light-cone are treated with the standard Wightman's prescription $t \to t - i\xi$, $\xi > 0$. 

It is instructive to check how the CME current comes out of (\ref{i1}), (\ref{i2}):
\be
J_3 = \lim\limits_{x\to x'} \T \gamma_3 \left( G_L(x,x') + G_R(x,x')\right)
\label{ss}
\ee
One can see that the free term (\ref{free}) does not contribute, while (\ref{b}) takes the form
\be
G^{(1)}(\mu,t,z) = -\frac{2}{t^2 - z^2} \left[ (\gamma_0 t - \gamma_3 z )  + i \mu z (\gamma_3 t - \gamma_0 z)  + {\cal O}(t,z) \right]
\ee
and provides\footnote{As usually, the coincidence limit is to be taken after the computation of the trace, not before.}
the correct answer in the limit $t,z \to 0$:
\be
J_3 = \frac{eB}{4\pi^2} (\mu_R - \mu_L)
\ee

We can now compute the main quantity of our interest  $ - \delta {\cal F}(\omega)$, the difference of the response functions corresponding to the currents parallel to the magnetic field to transverse to it. Roughly speaking, this quantity tells us which detector clicks more often (or, in other words, which one has more clicks for a given time interval): the detector measuring the currents along the field or the one oriented to measure the orthogonal currents. This difference is an integral of the following integrand:
$$
\delta G^+ (\tau - \tau')  = \lan \Omega_0 | j_3(x(\tau)) j_3(x(\tau'))  - j_1(x(\tau)) j_1(x(\tau'))| \Omega_0 \rrr =
$$
\be
=J_3^2 + \frac{(eB)^2}{8 \pi^4} \frac{1}{(\tau - \tau')^2}
\label{p4}
\ee
where $J_3$ is given by (\ref{ss}), and worldline of the detector is taken as $x(\tau) = (\tau,0,0,0)$ (detector at rest). Notice that there is no contribution from $G^{(0)} G^{(0)}$ and mixed $G^{(0)} G^{(1)}$ terms.

The first term corresponds to the usual CME current and due to its stationary nature it cannot be detected by stationary detector (i.e. for $f(\tau)$ identically equal to unity). The same is true for the second term, since \be
\int\limits_{-\infty}^{\infty} ds \> e^{-i\omega s} \> \frac{1}{(s-i\xi)^{2}} = 0
\label{ikj}
\ee
for $\omega > 0$. It this respect the situation at finite density is different from that at finite temperature. The thermal state has excitations of any desired energy (corresponding to poles along imaginary time axes in integral (\ref{eri})), which can excite the detector. The finite density state is a state of the lowest energy which has to be excited "by hands" (i.e. by nontrivial $f(\tau)$) to get observable results.

Inserting (\ref{p4}) into (\ref{or}) we get the following answer:
\be
\delta {\cal F} (\omega) = \frac{(eB)^2}{4\pi^4} \left[\mu_5^2 I_0 - \frac{1}{2} I_2\right]
\label{main}
\ee
where
\be
I_n = \int f(\tau) d\tau \int f(\tau') d\tau' e^{-i\omega (\tau - \tau')} \cdot \frac{1}{(\tau - \tau' - i \xi)^n}
\label{p7}
\ee
As in \cite{Shevchenko2}, the result (\ref{main}) is exact in $B$, i.e. the asymmetry $\delta {\cal F} (\omega)$ gets no contributions from higher powers of magnetic field.

The integral (\ref{p7}) depends on typical measurement time $\lambda$  encoded in $f(\tau)$ and also on dimensionless variables $\omega \lambda$ and $\xi /\lambda$. The infinitesimal parameter $\xi$ from Wightman's prescription has physical meaning of inverse maximal particle energy which can be measured by the detector and which is necessary a finite quantity for any realistic detector. From another point of view, it can be seen as related to the finite size of the detector \cite{FS}. The quantum measurement theory with finite measurement windows has physical sense and can be applicable only for $\lambda \gg \xi$. However to keep $\xi \neq 0$ is important to get formally correct $\lambda \to 0$ limit, i.e. when the detector is not switched on at all and should, correspondingly, return zero count (see discussion of this issue in \cite{Padm}).

The concrete form of (\ref{p7}) depends, of course, on a chosen detector switching time profile. In the problem in question it is natural to identify it with the time profile of the magnetic field itself. The latter was computed by many authors \cite{Skokov1,Skokov2,Toneev}. For example, it was suggested in \cite{Kharzeev10} to use the following Ansatz:
\be
B(\tau) = \frac{B_0}{(1+(\tau/\lambda)^2)^{3/2}}
\label{bi}
\ee
where $B_0$ and $\lambda$ are functions of impact parameter and rapidity, whose typical scale for RHIC setup is given by $B_0 \sim 10^5 \> \mbox{MeV}^2$, $\lambda \sim 0.1 \> \mbox{Fm}/c$. It is of course not quite correct simply  replace $B \to B(\tau)$ in (\ref{p4}) since the exact Green's function (\ref{i1}) is written for time-independent $B$. But since our approach anyway depends on a detector's model, we find it is a reasonable approximation to use (\ref{bi}) as a profile for the window function $f(\tau)$. The result for (\ref{p7}) with $n=0$ follows trivially:
\be
I_0 = 4 {\omega^2} \lambda^4 \>K_1^2(\kappa)
\ee
where $\kappa = \omega \lambda$. The maximum of this function in $\kappa$ (i.e. optimal measurement time) is reached at $\lambda \approx 1.33/\omega$. It is easy to see that both at $\lambda \to 0$ (no measurement at all) and $\lambda \to \infty$ (stationary measurement) $I_0$ vanishes.

The computation of $I_2$ is a little bit more tricky, the result reads:
\be
I_2 = - 4 \int\limits_{\kappa}^\infty dz (z-\kappa) z^2 {{K}}_1^2(z)\> e^{-\xi (z-\kappa)/\lambda}
\ee
Again both limits $\lambda \to 0 , \infty$ are respected. For finite $\lambda$ one can remove the cutoff and put $\xi = 0$.
Having these results we can rewrite (\ref{main}) as
\be
\delta {\cal F} (\omega) = \frac{(eB)^2}{4\pi^4} I_0 \left[\mu_5^2 + \frac{1}{\lambda^2} g(\kappa)\right]
\label{main2}
\ee
with the dimensionless function $g(\kappa)$ given by
\be
g(\kappa) = \frac{\kappa^2}{2} \int\limits_{1}^\infty dy y^2 (y-1) \left( \frac{K_1(y\kappa)}{K_1(\kappa)} \right)^2
\label{g3}
\ee
We have put $\xi = 0$, assuming finiteness of $\kappa$.

The expression in square brackets in (\ref{main2}) can be called effective axial chemical potential. The physical meaning of it is quite transparent: by energy-momentum uncertainty relation the finite observation time $\lambda$ makes quarks Fermi energies uncertain, Dirac sea becomes wavy, and these fluctuations provide the vector current excess along the magnetic field even if "bare" axial chemical potential is absent. While the concrete form of the functions $I_0$, $g(\kappa)$ depends, of course, on the chosen window function profile, the qualitative form of the result (\ref{main2}) is robust. In particular, the same conclusion about the structure of "effective" $\mu_5$ is reached in \cite{Shevchenko2} using decoherence functionals framework.

In quantitative terms, $g(\kappa \sim 1)\approx 0.15 \div 0.25$ and the model (\ref{main2}) with (\ref{g3}) corresponds to effective axial chemical potential of order of 1 GeV, even if "bare"  $\mu_5 = 0$. It is worth stressing that in the discussed picture the current excess resulted from $f(\tau) \neq 1$ is indistinguishable from the one caused by $\mu_5 \neq 0$.

Let us make a final remark. Usually in the problems of the discussed type one neglects any kind of back reaction, i.e. the external parameters (temperature, fields, acceleration etc) are taken as constants. However this condition can be relaxed - at least in adiabatic approximation it is easy to address the question of energy balance for the whole system including the bath and the detector. One can think of two limiting models  of energy release from the system: "fast" release and "slow" one. In the former case the excited detector takes away its excitation energy from the system in negligibly short time and gets back to the initial state. So for unit time interval the detector can be excited any number of times in this limit. The latter, "slow" release case corresponds to assumption that after excitation the detector stays in the final state, so no more than a unit quantum of excitation energy can be taken away from the bath by one detector for any finite time interval.

From the structure of finite density Green's functions (\ref{free}), (\ref{b}) it can be seen that for the detector to click in the course of stationary measurement at finite density, it should move with respect to laboratory rest frame with the speed $v$ (i.e. its worldline is to be $(t,0,0, \pm vt)$) such that $\omega - 2\mu v < 0$ (in which case the integrals like (\ref{ikj}) return nonzero answer even if $f(\tau) \equiv 1$). The physical meaning of this condition is simple - the energy gap between particles, comoving and anti-comoving with the detector should exceed the energy gap of the detector in its rest frame. In this case the detector (moving with zero acceleration!) will detect free finite density currents. It is worth repeating that the stationary detector at rest detects nothing at finite density because, in terms of (\ref{y7}), $| \Omega_0 \rrr$ is the stationary state of lowest energy.

Following this logic one can study, for example, the freezing of a thermal bath of some field by the Unruh-DeWitt detectors (at rest or moving) immersed into it. The same considerations apply to chiral current. The often cited statement about its nondissipative nature deals with the strictly stationary regime. However for time-dependent process (where time dependence may be induced by $B$ and/or by $\mu_5$ as well as by the act of measurement itself) this is no longer true, as expressions (\ref{main}), (\ref{main2}) clearly demonstrate. This should be of direct importance for the systems supporting chiral currents and coupled to "normal" electromagnetic devices, realizing "chiral electronics" circuits \cite{CE}. This interesting question deserves further study.

\section{P-odd effects in central collisions}

We have shown in the previous section that non-stationary nature of the heavy ion collision process could result in charge fluctuation asymmetry with respect to the reaction plane, and this asymmetry is of the same sign as one would expect for CME. The same qualitative result was obtained in \cite{Shevchenko2} for thermal fluctuations. Since in both cases the effect is related to abelian anomaly, one might say that CME {\it by definition is nothing but these asymmetries}. This would not be terminologically correct, however. In the original framework CME has been seen \cite{Kharzeev7} as a manifestation of {\it P-odd effects of nonabelian nature}. In some sense, the role of magnetic field in noncentral collisions is just to make P-odd chirality imbalance manifest by electric charge asymmetry of final particles, while the imbalance itself (encoded, e.g. in nonzero $\mu_5$) is supposed to be generated by some nonperturbative mechanism of QCD origin. In the former cases, on the contrary, free fermions have been considered. Roughly speaking, we demonstrated that one does not need to have nonzero $\mu_5$ to get CME-like charge asymmetry. This however does not at all mean, that there is no effective nonzero $\mu_5$ in real heavy-ion collision events and all the corresponding P-odd physics like CME (understood in its original way) related to that. In the light of the above it is worth to think (to rethink, see e.g. \cite {vv}) about possible P-odd signatures, {\it not related} to the magnetic field.

The following simple analogy could be useful. In order to measure the standard electric conductivity of some medium one can proceed along the following two ways. The first recipe is to apply external electric field ${\cal E}(\omega)$, measure induced electric current $J(\omega)$ and get the (complex) resistivity from the Ohm's law ${\cal E}(\omega) = Z(\omega)\cdot J(\omega)$. The second way is according to Nyquist - measure fluctuations (thermal and quantum) of electric current in the medium {\it without} any external field and extract conductivity from
\be
\left(J^2\right)_\omega = \hbar \omega \frac{R(\omega)}{|Z(\omega)|^2}  \cot\frac{\hbar
\omega}{2kT}
\ee
where $R(\omega) = \Re{\{ Z(\omega) \} }$.The relation between the above two approaches is of course the essence of fluctuation-dissipation theorem \cite{Callen}.
Despite the chiral conductivity is T-even and the corresponding current is non-dissipative, general linear response theory arguments are applicable here as well.
Indeed, one has for the induced current in linear response approximation \cite{Kharzeev10}:
\be
\lan j_\mu(x) \rrr = \int d^4 x' \> \Pi_{\mu\nu}^{(R)}(x,x') A^\nu(x')
\ee
where the retarded polarization operator is given by
\be
\Pi_{\mu\nu}^{(R)}(x,x') = i \theta(x_0-x'_0) \lan [j_\mu(x), j_\nu(x')] \rrr
\label{pl}
\ee
Assuming there is a distinguished rest frame, parameterized by constant vector $u_\mu$, normalized to unity $u_\mu u^\mu = 1$, the Fourier transform of (\ref{pl}) can be decomposed as follows (we are using tilde sign to distinguish Fourier components in what follows):
\be
{\tilde\Pi}_{\mu\nu}^{(R)}(q,u) = \sum\limits_{i=1}^3 \>\Psi^{(i)}_{\mu\nu} \>{\tilde\Pi}_i(q^2,qu)
\label{decomp}
\ee
with the following tensor structures:
$$
\begin{array}{l}
\Psi^{(1)}_{\mu\nu} = g_{\mu\nu} q^2 - q_\mu q_\nu \\
\Psi^{(2)}_{\mu\nu}  = \left((qu) q_\mu - q^2 u_\mu \right)\left((qu) q_\nu - q^2 u_\nu \right) \\
\Psi^{(3)}_{\mu\nu} = i \epsilon_{\mu\nu\alpha\beta}q^\alpha u^\beta\\
\end{array}
$$
The decomposition (\ref{decomp}) respects current conservation $q^\mu {\tilde\Pi}_{\mu\nu}^{(R)}(q,u) =0$ and Bose-symmetry ${\tilde\Pi}_{\mu\nu}^{(R)}(q,u) = {\tilde\Pi}_{\nu\mu}^{(R)}(-q,u)$. The electric chiral current corresponds to the third term of this decomposition:
\be
\lan j_\mu(x) \rrr_\chi = \int \frac{d^4 q}{(2\pi)^4} e^{-iqx} \> {\tilde\Pi}^{(3)}(q^2, qu) u^\beta {\tilde F}_{\mu\beta}(q)
\ee
where
\be
{\tilde F}_{\mu\beta}(q) = \frac12 \epsilon_{\mu\beta\nu\alpha} \int d^4 x \> e^{iqx} \> F^{\nu\alpha}(x)
\ee
For free fermions with chemical potentials $\mu_L$, $\mu_R$ the chiral conductivity is exactly computed in this way in \cite{Kharzeev10} and in static limit coincides with (1), as it should be:
\be
 \lim\limits_{q\to 0} \Pi^{(3)}(q^2, qu) = \frac{e\mu_5}{2\pi^2}
\ee

Without loss of generality one can take $u_\mu=(1,0,0,0)$. To extract a part proportional to the invariant function $\Pi^{(3)}$ it is convenient to rewrite the definition (\ref{pl}) for particular spatial components of the commutator $i=1$, $k=2$ and to choose $(x-x') = (\tau,0,0,z)$ with $\tau > 0$. The result reads:
\be
\lan [j_1(\tau,z/2), j_2(0,-z/2) ] \rrr = \int \frac{d^4 q}{(2\pi)^4} \> e^{-i(q_0 \tau - q_3 z)} q_3 \> {\tilde\Pi}^{(3)} (q^2,qu)
\label{kj}
\ee
Of course, any $\mathbb{O}(3)$-rotated choice of indices instead of (\ref{kj}) is equally legitimate.

The formula (\ref{kj}) expresses correlation of current fluctuations of some particular form in terms of the transport coefficient of interest. One can check that both sides of this expression change sign for $z\to -z$ and vanish at $z=0$. The skew-symmetric form of the l.h.s of (\ref{kj}) corresponds to non-dissipative nature of these correlations. It is also convenient to rewrite (\ref{kj}) in coordinate form:
\be
\Pi^{(3)}(\tau,\tau) - \Pi^{(3)}(\tau,0) = i \int\limits_0^\tau dz \lan [j_1(\tau,z), j_2(0,0)]\rrr
\label{kj1}
\ee

To get possible intuitive physical picture behind (\ref{kj}), imagine outgoing radial flow of massless fermions, all of the same chirality, emitted from the origin and absorbed ("detected") by a rigid spherical shell. For uniform flow total angular momentum of the shell after interaction is equal to zero by symmetry. Suppose now that the shell is cut along its equator.\footnote{In terms of (\ref{kj}) that equator plane has $z=0$. Notice that due to spherical symmetry of the problem the choice of equator is arbitrary.} Then after interaction both hemispheres start to rotate with angular momenta equal in magnitude and opposite in sign. It is obvious that for chirally balanced matter (with $\mu_5 = 0$) there is no such effect. Observation of the hemispheres rotation in such experiment would be a direct sign of P-odd physics behind. All that is in some correspondence with such phenomena as Chiral Magnetic Spiral \cite{CMS} and Chiral Magnetic Wave \cite{CMW} but there is no magnetic field in the discussed example.

It would be tempting to try {\it to measure} the r.h.s. of (\ref{kj1}) in a given heavy ion collision event and hence be able {\it to compute} $\Pi^{(3)}$ without any reference to non-centrality and magnetic field (i.e. to proceed in Nyquistic way, not in Ohmic one). However one could not expect simple correspondence between expressions like (\ref{kj1}) containing quantum commutators and final particles distributions which are essentially classical.\footnote{Like, for example, measuring commutator of $x$- and $y$-components of angular momentum $\lan [L_x , L_y] \rrr$ has to do with measuring $L_z $, but not $L_x$ and $L_y$.}

%
%
%

As is well known the final particle distribution in any heavy-ion collision event is conventionally parameterized as
$dN(y, p_\bot , \phi)$ where $y$ is rapidity $y=(\log(E+p_\|) - \log(E - p_\|))/2$, $\phi$ - azimuthal angle and $p_\bot$, $p_\|$ - transverse and longitudinal momenta, respectively. P-inverse of this fluctuation map can be obtained by the replacement $y \to -y$, $\phi \to \pi + \phi$ and in general it does not, of course, coincide with the original distribution $dN$ even for spherical nuclei experiencing exact central collision because of quantum, thermal and combinatorial fluctuations, which present at all stages starting from initial state of colliding nuclei till the freeze-out phase. One can introduce positive definite distance between the original and P-inverted distributions and study it as a function of collision's parameters (energy, type of nuclei etc). The question to answer is whether also {\it additional} source of P-odd asymmetry, related to internal strong interaction dynamics, is present, {\it besides} all mentioned above statistical factors.

Positive answer would mean the existence of some regular patterns in $dN$ (in particular, in central events), improbable for statistical fluctuations. The simple model of free fermions absorbed by the rigid hemispheres discussed above is just an example of such structure (i.e. nonzero average angular momentum carried by particles emitted in any given solid angle). We believe that searches for complementary observables, sensitive to nonperturbative P-odd effects at heavy ion collisions which are not related to existence of the magnetic field is extremely important to have a satisfactory theoretical picture of this interesting field of physics.

\bigskip

{\bf \Large Acknowledgements}

\medskip

This work began during the Workshop on QCD in strong magnetic fields in Italy, November 12-16, 2012, hosted by ECT in Trento. The author is grateful to the organizers of this workshop for inviting and stimulating atmosphere.

Discussions of various  CME-related topics with M.Chernodub, D.Kharzeev, V.Kirillin, M.Polikarpov, A.Sadofyev, I.Selyuzhenkov, O.Teryaev, S.Voloshin, V.Zakharov are acknowledged.

\newpage

\section{Appendix}

Despite theoretical framework for fermions at finite density is well developed \cite{Chodos}, we find it instructive to remind the basic steps here. First, we put external field to zero and  take free massless noninteracting fermions with chemical potentials $\mu_L$ and $\mu_R$ for chiral components. The system's Hamiltonian gets shifted by the terms:
\be
\Delta H = \mu_L \int d^3 {\bf x} \>{\bar\psi}_L \gamma_0 {\psi}_L + \mu_R \int d^3 {\bf x} \> {\bar\psi}_R \gamma_0 {\psi}_R \ee
where ${\psi}_L = (1-\gamma^5)\psi/2$, ${\psi}_R = (1+\gamma^5)\psi/2$. We assume in this paper that both chemical potentials are positive, corresponding to nonzero density of particles (and not antiparticles).

The corresponding modified dispersion relations are given by
\be
(p_0 - \mu_{L})^2 = {\bf p}^2 \;\;\; ; \;\; (p_0 - \mu_{R})^2 = {\bf p}^2
\label{dr}
\ee
for left and right chiral components, respectively. For chirally symmetric case $\mu_{L} = \mu_{R} = \mu$.

To simplify notation, we make for the moment no distinction between chiral modes. The key point, which makes (\ref{dr}) different from just unobservable shift of all energy levels is the choice of integration contour in the Green's function, which takes into account the fact that states between 0 and $\mu$ are occupied.\footnote{One can say, that the Dirac sea level is at $\mu$, not at zero.} It is fixed by the following prescription
\be
i\epsilon(p_0) = \left\{
\begin{array}{cc}
+i\epsilon , & \mu < p_0  \\
-i\epsilon , & 0<p_0 < \mu \\
+i\epsilon , & p_0 < 0 \\
\end{array}
\right.
\ee
Constant and uniform magnetic field is included in the standard way via covariant derivative: $(p_0, {\bf p}) \to (\pi_0, {\bf \pi} )$, where $\pi_0 = p_0 - \mu$, ${\bf{\pi}} = {\bf p} - e {\bf A}$ and ${\bf B} = \mbox{rot}\> {\bf A}$.
With these definitions, the $T$-ordered Green's function at finite density in external magnetic field is given by (in the massless case):
\be
G(x,x') = \lan \Omega | T\{ \psi(x) \bar\psi(x') \} | \Omega \rrr
= i \int \frac{d^4 p}{(2\pi)^4} \> \frac{\gamma \pi
}{(\gamma \pi)^2 +i\epsilon(p_0)} \> e^{-ip(x-x')}
\ee
It is convenient to use proper-time representation (\cite{Chodos}):
 \be
G(p) = \theta(\pi_0 \> {\mbox{sign}}\> p_0 )\> {\tilde S}(p) - \theta(-\pi_0 \> {\mbox{sign}}\> p_0)\>{\tilde S}^\dagger(p)
\ee
where the exact fermion propagator is given by \cite{tsai}:
\be
{\tilde S}(p) = \int\limits_0^\infty \> du \> e^{iu(\pi_0^2 - p_3^2 - p_\bot^2 \frac{\tan(eBu)}{eBu} + i\epsilon )} \left[ P_0 + P_1  \right]
\label{hf}
\ee
with
\be
P_0 = \pi_0 \gamma_0 - p_3 \gamma_3 - p_\bot \gamma_\bot (1+\tan^2(eBu))
\ee
\be
P_1 = (\pi_0 \gamma_0 - p_3 \gamma_3)\gamma_1 \gamma_2 \tan(eBu)
 \ee
and $p_\bot \gamma_\bot = p_1\gamma_1 + p_2 \gamma_2 $.
Another useful representation follows from the identity $\theta(z) + \theta(-z) =1$:
\be
G(p) = {\tilde S}(p) - \theta(-\pi_0 \> {\mbox{sign}}\> p_0)\>\left({\tilde S}(p) + {\tilde S}^\dagger(p)\right)
\label{ga}
\ee
In free case using the relation
\be
\frac{1}{x-i\epsilon} = P\> \frac{1}{x} + i\pi \delta(x)
\ee
it is easy to obtain for the Fourier transform of (\ref{ga})
\be
G_0(x) = {\tilde S}_0(x) -2\pi \int\frac{d^3{\bf p}}{(2\pi)^3} \int\limits_0^\mu \frac{dp_0}{2\pi} e^{-ipx} (\pi_0 \gamma_0 - {\bf p \gamma}) \delta(\pi_0^2 - {\bf p}^2)
\label{fg}
\ee
where ${\tilde S}_0(x) $ is Fourier transform of free propagator without external field:
\be
{\tilde S}_0(p) =
 \int\limits_0^\infty \> du \> e^{iu(\pi_0^2 - {\bf p}^2 + i\epsilon) }
\ee

Only the second term in the r.h.s. of (\ref{fg}) contributes to density $\lan \psi^\dagger \psi \rrr$ and other on-shell quantities.

With the external magnetic field it is not possible to get on-shell term in the same simple representation as in (\ref{fg}). The situation however drastically simplifies for the case of zero transverse distance $(x-x')_\bot = 0$. This is exactly the kinematics we are interested in. One has:
$$
\int \frac{d^2 p_\bot}{(2\pi)^2} \left( S(p) + S^\dagger(p) \right) = (\pi_0 \gamma_0 - p_3\gamma_3) \left[ \int \frac{d^2 p_\bot}{2\pi}  \delta (\pi_0^2 - {\bf p}^2) + \right.
$$
\be
+ \left.  \frac{1}{2\pi} \int\limits_0^{\infty} e^{-\epsilon u} \sin u (\pi_0^2 - p_3^2) \left( \frac{eB}{\tan eBu} - \frac{1}{u} \right)  +
\frac{\gamma_1 \gamma_2 eB }{2 i} \delta(\pi_0^2 - p_3^2) \right]
\label{fg2}
\ee
Performing Fourier transformation of (\ref{fg}) and taking into account (\ref{fg2}) one obtains the expressions for Green's functions used in the main text.


\begin{thebibliography}{199}



\bibitem{Voloshin1}
  S.~A.~Voloshin,
  ``Discussing the possibility of observation of parity violation in heavy  ion
  collisions,''
  Phys.\ Rev.\  C {\bf 62}, 044901 (2000)
  [arXiv:nucl-th/0004042].


\bibitem{Voloshin2}
  S.~A.~Voloshin,
  ``Parity violation in hot QCD: How to detect it,''
  Phys.\ Rev.\  C {\bf 70}, 057901 (2004)
  [arXiv:hep-ph/0406311].




\bibitem{Seluzhenkov}
  I.~V.~Selyuzhenkov  [STAR Collaboration],
  ``Global polarization and parity violation study in Au + Au collisions,''
  Rom.\ Rep.\ Phys.\  {\bf 58}, 049 (2006)
  [arXiv:nucl-ex/0510069].


\bibitem{Selyuzhenkov2}
  I.~Selyuzhenkov  [STAR Collaboration],
  ``Azimuthal charged particle correlations as a probe for local strong parity
  violation in heavy-ion collisions,''
  arXiv:0910.0464 [nucl-ex].


\bibitem{Voloshin7}
  S.~A.~Voloshin,
  ``Local strong parity violation and new possibilities in experimental study
  of non-perturbative QCD,''
  arXiv:1003.1127 [nucl-ex].



\bibitem{Voloshin3}
  S.~A.~Voloshin,
  ``Anisotropic flow: Achievements, Difficulties, Expectations,''
  J.\ Phys.\ G {\bf 35}, 104014 (2008)
  [arXiv:0805.1351 [nucl-ex]].

\bibitem{Voloshin4}
  S.~A.~Voloshin  [STAR Collaboration],
  ``Probe for the strong parity violation effects at RHIC with three particle
  correlations,''
  arXiv:0806.0029 [nucl-ex].


\bibitem{Voloshin5}
  S.~A.~Voloshin,
  ``Anisotropic collective phenomena in ultra-relativistic nuclear
  collisions,''
  Nucl.\ Phys.\  A {\bf 827}, 377C (2009)
  [arXiv:0902.0581 [nucl-ex]].


\bibitem{Voloshin6}
  S.~A.~Voloshin  [STAR Collaboration],
  ``Experimental study of local strong parity violation in relativistic nuclear
  collisions,''
  Nucl.\ Phys.\  A {\bf 830}, 377C (2009)
  [arXiv:0907.2213 [nucl-ex]].


\bibitem{Abelev1}
  B.~I.~Abelev {\it et al.}  [STAR Collaboration],
  ``Observation of charge-dependent azimuthal correlations and possible local
  strong parity violation in heavy ion collisions,''
  arXiv:0909.1717 [nucl-ex].

\bibitem{Abelev2}
  B.~I.~Abelev {\it et al.}  [STAR Collaboration],
  ``Azimuthal Charged-Particle Correlations and Possible Local Strong Parity
  Violation,''
  Phys.\ Rev.\ Lett.\  {\bf 103}, 251601 (2009)
  [arXiv:0909.1739 [nucl-ex]].


\bibitem{Wang}
  G.~Wang  [STAR Collaboration],
  ``Highlights from STAR: probing the early medium in heavy ion collisions,''
  Nucl.\ Phys.\  A {\bf 830}, 19C (2009)
  [arXiv:0907.4504 [nucl-ex]].

\bibitem{Kharzeev1}
  D.~Kharzeev, R.~D.~Pisarski and M.~H.~G.~Tytgat,
  ``Possibility of spontaneous parity violation in hot {QCD},''
  Phys.\ Rev.\ Lett.\  {\bf 81}, 512 (1998)
  [arXiv:hep-ph/9804221].


\bibitem{Kharzeev2}
  D.~Kharzeev and R.~D.~Pisarski,
  ``Pionic measures of parity and CP violation in high energy nuclear
  collisions,''
  Phys.\ Rev.\  D {\bf 61}, 111901 (2000)
  [arXiv:hep-ph/9906401].


\bibitem{Kharzeev3}
  D.~E.~Kharzeev, R.~D.~Pisarski and M.~H.~G.~Tytgat,
  ``Aspects of parity, CP, and time reversal violation in hot QCD,''
  arXiv:hep-ph/0012012.


\bibitem{Kharzeev4}
  D.~Kharzeev, A.~Krasnitz and R.~Venugopalan,
  ``Anomalous chirality fluctuations in the initial stage of heavy ion
  collisions and parity odd bubbles,''
  Phys.\ Lett.\  B {\bf 545}, 298 (2002)
  [arXiv:hep-ph/0109253].


\bibitem{Kharzeev5}
  D.~Kharzeev,
  ``Parity violation in hot QCD: Why it can happen, and how to look for it,''
  Phys.\ Lett.\  B {\bf 633}, 260 (2006)
  [arXiv:hep-ph/0406125].



\bibitem{Kharzeev6}
  D.~Kharzeev and A.~Zhitnitsky,
  ``Charge separation induced by P-odd bubbles in QCD matter,''
  Nucl.\ Phys.\  A {\bf 797}, 67 (2007)
  [arXiv:0706.1026 [hep-ph]].


\bibitem{Kharzeev7}
  D.~E.~Kharzeev, L.~D.~McLerran and H.~J.~Warringa,
  ``The effects of topological charge change in heavy ion collisions: 'Event by
  event P and CP violation',''
  Nucl.\ Phys.\  A {\bf 803}, 227 (2008)
  [arXiv:0711.0950 [hep-ph]].



\bibitem{Kharzeev8}
  D.~E.~Kharzeev,
  ``Hot and dense matter: from RHIC to LHC: Theoretical overview,''
  Nucl.\ Phys.\  A {\bf 827}, 118C (2009)
  [arXiv:0902.2749 [hep-ph]].

\bibitem{Kharzeev9}
  D.~E.~Kharzeev,
  ``Topologically induced local P and CP violation in hot QCD,''
  arXiv:0906.2808 [hep-ph].


\bibitem{Kharzeev10}
  D.~E.~Kharzeev and H.~J.~Warringa,
  ``Chiral Magnetic conductivity,''
  Phys.\ Rev.\  D {\bf 80}, 034028 (2009)
  [arXiv:0907.5007 [hep-ph]].

\bibitem{Kharzeev11}
  D.~E.~Kharzeev,
  ``Chern-Simons current and local parity violation in hot QCD matter,''
  Nucl.\ Phys.\  A {\bf 830}, 543C (2009)
  [arXiv:0908.0314 [hep-ph]].


\bibitem{Fukushima1}
  K.~Fukushima, D.~E.~Kharzeev and H.~J.~Warringa,
  ``The Chiral Magnetic Effect,''
  Phys.\ Rev.\  D {\bf 78}, 074033 (2008)
  [arXiv:0808.3382 [hep-ph]].



\bibitem{Warringa1}
  H.~J.~Warringa,
  ``Implications of CP-violating transitions in hot quark matter on heavy ion
  collisions,''
  J.\ Phys.\ G {\bf 35}, 104012 (2008)
  [arXiv:0805.1384 [hep-ph]].




\bibitem{Warringa2}
  H.~J.~Warringa,
  ``The Chiral Magnetic Effect: Measuring event-by-event P- and CP-violation
  with heavy ion-collisions,''
  arXiv:0906.2803 [hep-ph].




\bibitem{Fraga}
  E.~S.~Fraga and A.~J.~Mizher,
  ``Chiral symmetry restoration and strong CP violation in a strong magnetic
  background,''
  PoS C {\bf POD2009}, 037 (2009)
  [arXiv:0910.4525 [hep-ph]].


\bibitem{Nam1}
  S.~i.~Nam,
  ``Chiral magnetic effect at low temperature,''
  Phys.\ Rev.\  D {\bf 80}, 114025 (2009)
  [arXiv:0911.0509 [hep-ph]].

\bibitem{Abramczyk}
  M.~Abramczyk, T.~Blum, G.~Petropoulos and R.~Zhou,
  ``Chiral magnetic effect in 2+1 flavor QCD+QED,''
  arXiv:0911.1348 [hep-lat].



\bibitem{McLerran}
  L.~McLerran,
  ``Theoretical Concepts for Ultra-Relativistic Heavy Ion Collisions,''
  arXiv:0911.2987 [hep-ph].

\bibitem{Kharzeev12}
  D.~E.~Kharzeev,
  ``Topologically induced local P and CP violation in QCD x QED,''
  Annals Phys.\  {\bf 325}, 205 (2010)
  [arXiv:0911.3715 [hep-ph]].



\bibitem{Nam2}
  S.~i.~Nam,
  ``Chiral magnetic effect (CME) at low temperature from instanton vacuum,''
  arXiv:0912.1933 [hep-ph].

\bibitem{Fukushima2}
  K.~Fukushima, D.~E.~Kharzeev and H.~J.~Warringa,
  ``Electric-current Susceptibility and the Chiral Magnetic Effect,''
  arXiv:0912.2961 [hep-ph].



\bibitem{Bzdak}
  A.~Bzdak, V.~Koch and J.~Liao,
  ``Remarks on possible local parity violation in heavy ion collisions,''
  arXiv:0912.5050 [nucl-th].


\bibitem{Fu}
  W.~j.~Fu, Y.~x.~Liu and Y.~l.~Wu,
  ``Chiral Magnetic Effect and Chiral Phase Transition,''
  arXiv:1002.0418 [hep-ph].



\bibitem{Fukushima3}
  K.~Fukushima, D.~E.~Kharzeev and H.~J.~Warringa,
  ``Real-time dynamics of the Chiral Magnetic Effect,''
  arXiv:1002.2495 [hep-ph].





\bibitem{Buividovich1}
  P.~V.~Buividovich, M.~N.~Chernodub, E.~V.~Luschevskaya and M.~I.~Polikarpov,
  ``Chiral magnetization of non-Abelian vacuum: a lattice study,''
  Nucl.\ Phys.\  B {\bf 826}, 313 (2010)
  [arXiv:0906.0488 [hep-lat]].



\bibitem{Buividovich2}
  P.~V.~Buividovich, M.~N.~Chernodub, E.~V.~Luschevskaya and M.~I.~Polikarpov,
  ``Numerical evidence of chiral magnetic effect in lattice gauge theory,''
  Phys.\ Rev.\  D {\bf 80}, 054503 (2009)
  [arXiv:0907.0494 [hep-lat]].






\bibitem{Buividovich3}
  P.~V.~Buividovich, M.~N.~Chernodub, E.~V.~Luschevskaya and M.~I.~Polikarpov,
  ``Lattice QCD in strong magnetic fields,''
  arXiv:0909.1808 [hep-ph].



\bibitem{Buividovich4}
  P.~V.~Buividovich, M.~N.~Chernodub, E.~V.~Luschevskaya and M.~I.~Polikarpov,
  ``Numerical study of chiral magnetic effect in quenched SU(2) lattice gauge
  theory,''
  arXiv:0910.4682 [hep-lat].




\bibitem{Shevchenko1}
 V.~D.~Orlovsky and V.~I.~Shevchenko,
  ``Towards a quantum theory of chiral magnetic effect,''
  Phys.\ Rev.\ D {\bf 82} (2010) 094032
  [arXiv:1008.4977 [hep-ph]].

\bibitem{Shevchenko2}
V.~Shevchenko,
  ``Quantum measurements and chiral magnetic effect,''
  Nucl.\ Phys.\ B {\bf 870} (2013) 1
  [arXiv:1208.0777 [hep-th]].

\bibitem{mcl}

 E.~Iancu, A.~Leonidov and L.~D.~McLerran,
  ``Nonlinear gluon evolution in the color glass condensate. 1.,''
  Nucl.\ Phys.\ A {\bf 692} (2001) 583
  [hep-ph/0011241].

\bibitem{Pomeranchuk}
 I.~Y.~.Pomeranchuk,
  ``On the theory of multiple particle production in a single collision,''
  Dokl.\ Akad.\ Nauk Ser.\ Fiz.\  {\bf 78} (1951) 889.

\bibitem{Unruh}
W.~G.~Unruh,
  ``Notes on black hole evaporation,''
  Phys.\ Rev.\ D {\bf 14} (1976) 870.

\bibitem{Zubkov}
M.I.Katsnelson, G.E.Volovik, M.A.Zubkov, 
``Unruh effect in vacua with anisotropic scaling: Applications to multilayer graphene,''
Annals of Physics {\bf 336} (2013) 36

\bibitem{Braginsky}
V.Braginsky, F. Khalili, {\it Quantum Measurement,} Cambridge University Press, 1992.

\bibitem{Birrel}
N. D. Birrell, P. C. W. Davies, 
{\it Quantum Fields in Curved Space}, Cambridge University Press, 1984

\bibitem{Bessa}
 C.~H.~G.~Bessa, J.~G.~Duenas and N.~F.~Svaiter,
``Accelerated detectors in Dirac vacuum: the effects of horizon fluctuations,''
  Class.\ Quant.\ Grav.\  {\bf 29} (2012) 215011
  [arXiv:1204.0022 [hep-th]].


\bibitem{Padm}
 L.~Sriramkumar and T.~Padmanabhan,
  ``Response of finite time particle detectors in noninertial frames and curved space-time,''
  Class.\ Quant.\ Grav.\  {\bf 13} (1996) 2061
  [gr-qc/9408037].

\bibitem{Chodos}
 A.~Chodos, K.~Everding and D.~A.~Owen,
  ``QED With A Chemical Potential: 1. The Case Of A Constant Magnetic Field,''
  Phys.\ Rev.\  D {\bf 42}, 2881 (1990).

\bibitem{Shuryak}
 E.~V.~Shuryak,
  ``Quantum Chromodynamics and the Theory of Superdense Matter,''
  Phys.\ Rept.\  {\bf 61} (1980) 71.


\bibitem{FS}
S.~Schlicht,
  ``Considerations on the Unruh effect: Causality and regularization,''
  Class.\ Quant.\ Grav.\  {\bf 21} (2004) 4647
  [gr-qc/0306022].



\bibitem{Skokov1}
 A.~Bzdak and V.~Skokov,
  ``Event-by-event fluctuations of magnetic and electric fields in heavy ion collisions,''
  Phys.\ Lett.\ B {\bf 710} (2012) 171
  [arXiv:1111.1949 [hep-ph]].

\bibitem{Skokov2}
V.~Skokov, A.~Y.~.Illarionov and V.~Toneev,
  ``Estimate of the magnetic field strength in heavy-ion collisions,''
  Int.\ J.\ Mod.\ Phys.\ A {\bf 24} (2009) 5925
  [arXiv:0907.1396 [nucl-th]].

  \bibitem{Toneev}
    V.~P.~Konchakovski, V.~Voronyuk, V.~D.~Toneev, W.~Cassing, E.~L.~Bratkovskaya and S.~A.~Voloshin,
  ``Evolution of electro-magnetic fields in relativistic heavy-ion collisions from the HSD transport approach,''
  PoS WPCF {\bf 2011} (2011) 045.


\bibitem{CE}
 D.~E.~Kharzeev and H.~-U.~Yee,
  ``Chiral Electronics,''
  arXiv:1207.0477 [cond-mat.mes-hall].


\bibitem{vv}
 S.~A.~Voloshin,
  ``Discussing the possibility of observation of parity violation in heavy ion collisions,''
  Phys.\ Rev.\ C {\bf 62} (2000) 044901
  [nucl-th/0004042].

\bibitem{Callen}
H.~B.~Callen and T.~A.~Welton,
  ``Irreversibility and generalized noise,''
  Phys.\ Rev.\  {\bf 83} (1951) 34.

\bibitem{CMS}
 G.~Basar, G.~V.~Dunne and D.~E.~Kharzeev,
  ``Chiral Magnetic Spiral,''
  Phys.\ Rev.\ Lett.\  {\bf 104} (2010) 232301
  [arXiv:1003.3464 [hep-ph]].

\bibitem{CMW}
 D.~E.~Kharzeev and H.~-U.~Yee,
  ``Chiral Magnetic Wave,''
  Phys.\ Rev.\ D {\bf 83} (2011) 085007
  [arXiv:1012.6026 [hep-th]].


\bibitem{tsai}
W.~-y.~Tsai,
  ``Modified electron Propagation Function in Strong Magnetic Fields,''
  Phys.\ Rev.\ D {\bf 10} (1974) 1342.



\end{thebibliography}
\end{document}